\documentclass[aps,pra,showpacs,twocolumn,superscriptaddress]{revtex4-1}
\usepackage{amsmath}
\usepackage{amssymb}
\usepackage{graphicx}
\usepackage{epstopdf}
\usepackage{times,txfonts}
\usepackage{easybmat}
\usepackage{mathtools}
\newcommand{\ket}[1]{|#1\rangle}
\newcommand{\bra}[1]{\langle #1|}
\usepackage[bookmarks=false]{hyperref}
\hypersetup{colorlinks=true,citecolor=blue,linkcolor=blue,urlcolor=blue,pdfstartview=FitH,bookmarksopen=true}
\usepackage{color,soul}

\begin{document}
\title{Rydberg-atom-based scheme of nonadiabatic geometric quantum computation}
\author{P. Z.  Zhao}
\affiliation{Department of Physics, Shandong University, Jinan 250100, China}
\author{Xiao-Dan Cui}
\affiliation{Department of Physics, Shandong University, Jinan 250100, China}
\author{G. F. Xu}
\email{sduxgf@163.com}
\affiliation{Department of Physics, Shandong University, Jinan 250100, China}
\affiliation{Department of Physics and Astronomy, Uppsala University,
Box 516, Se-751 20 Uppsala, Sweden}
\author{Erik Sj\"oqvist}
\email{erik.sjoqvist@physics.uu.se}
\affiliation{Department of Physics and Astronomy, Uppsala University,
Box 516, Se-751 20 Uppsala, Sweden}
\author{D. M. Tong}
\email{tdm@sdu.edu.cn}
\affiliation{Department of Physics, Shandong University, Jinan 250100, China}
\date{\today}

\begin{abstract}
Nonadiabatic geometric quantum computation provides a means to perform fast and robust quantum gates.
It has been implemented in various physical systems, such as trapped ions, nuclear magnetic resonance and superconducting circuits. Another system being adequate for implementation of nonadiabatic geometric quantum computation may be Rydberg atoms, since their internal states have very long coherence time and the Rydberg-mediated interaction facilitates the implementation of a two-qubit gate. Here, we propose a scheme of nonadiabatic geometric quantum computation based on Rydberg atoms, which combines the robustness of nonadiabatic geometric gates with the merits of Rydberg atoms.
\end{abstract}

\maketitle

\section{Introduction}

Quantum computation is believed much more effective than classical computation in solving many problems, such as factoring large integers \cite{Shor} and searching unsorted data \cite{Grover}.  The key for implementing circuit-based quantum computation is to realize a universal set of quantum gates with high fidelities. However, the errors arising from inaccurate control of a quantum system inevitably affect quantum gates, and the accumulation of control errors may seriously spoil the practical realization. This motivates researchers to utilize the characteristic of geometric phases to suppress control errors. Geometric phases are only dependent on evolution paths but independent of evolution details so that quantum computation based on geometric phases is robust against control errors.

The early schemes of geometric quantum computation \cite{Jones,Zanardi,Duan} were based on adiabatic geometric phases, both Abelian \cite{Berry} and non-Abelian phases \cite{Wilczek}. While adiabatic geometric gates are robust against control errors, an unavoidable challenge is the long run-time needed by adiabatic evolution, which makes the gates vulnerable to environment-induced decoherence and thereby hinders the experimental implementation. To overcome this difficulty, quantum computation \cite{Wang,Zhu,Sjoqvist2012,Xu2012} based on nonadiabatic Abelian \cite{Aharonov} and non-Abelian \cite{Anandan} geometric phases was put forward. Nonadiabatic geometric quantum computation has the merits of both high-speed implementation and robustness against control errors, and therefore it has received increasing attention \cite{Zhu2003,Zhu2003PRA,Solinas2003,Zhang2005,Feng2007,Feng2009,Kim2008,Thomas2011,Mousolou2014PRA,
Xu2014,Xu2014PRA,Mousolou2014NJP,Liang2014,Zhou2015,Xue2015,Xue2016,You2016,Xue2017,Xu2015,
Sjovist2016,Sjovist2016PRA,Zhao2016,Zhao2017,Xu2017,Xue2017PRA,Leibfried2003,
Du2006,Long,Abdumalikov,Arroyo,Duan2014}. A number of schemes for its physical implementation have been put forward \cite{Zhu2003PRA,Solinas2003,Zhang2005,Feng2007,Feng2009,Kim2008,Liang2014,Zhou2015,Xue2015,Xue2016,You2016,Xue2017},
and nonadiabatic geometric quantum computation has been experimentally demonstrated with trapped ions \cite{Leibfried2003}, NMR \cite{Du2006,Long}, superconducting circuits \cite{Abdumalikov} and nitrogen-vacancy centers in diamond \cite{Arroyo,Duan2014}.

As a promising candidate for implementing quantum computation, Rydberg atoms seem particularly attractive due to the long coherence time of internal atomic states and the realistically controllable Rydberg-mediated interaction between Rydberg atoms \cite{Saffman2005,Saffman2010}. The long-lived Rydberg states can be taken as well-defined qubit states, and the Rydberg-mediated interaction can lead to Rydberg blockade regime \cite{Jaksch2000,Lukin2001}, which allows an effective implementation of two-qubit gates. The Rydberg-mediated dipole-dipole or van der Waals interaction between high-lying Rydberg states is strong enough to shift atomic energy levels. It leads to Rydberg blockade, which prevents more than one atom from being excited to Rydberg states by resonant laser pulses. Rydberg blockade has been experimentally demonstrated with individual atoms \cite{Urban2009, G2009} as well as mesoscopic atomic ensembles \cite{Tong2004,Singer2004,Liebisch2005,Heidemann2007,Ebert2015,Weber2015}, and has been extensively applied to quantum computation \cite{Moller2008,Muller2009,Isenhower2010,Wu2010,Rao2014,Muller2014,Keating2015,Maller2015,
Sarkany2015,Barato2014,He2014,Khazali2015,Das2016}, including adiabatic geometric quantum computation \cite{Moller2008}. However, nonadiabatic geometric quantum computation based on Rydberg atoms has not been developed yet.

In this paper, we propose a Rydberg atom-based scheme of nonadiabatic geometric quantum computation, of which the computational qubit is encoded into the stable ground state and long-lived Rydberg state.  The one-qubit gates are performed by addressing an individual atom with laser pulse, and a nontrivial two-qubit gate is realized with the aid of the Rydberg blockade regime.
Our scheme combines the robustness and the speediness of nonadiabatic geometric gates with the merits of Rydberg atoms, and thereby provides a promising way of high-fidelity quantum computation.

The paper is organized as follows. In Sec. II, an arbitrary one-qubit nonadiabatic geometric gate  is proposed, which is based on Rydberg atoms but also applicable to other two-level systems. In Sec. III, a nontrivial two-qubit nonadiabatic geometric gate based on Rydberg atoms is proposed. In Sec. IV, we briefly discuss the feasibility of our proposal.  Section V is our conclusion.

\section{One-qubit gates}

\begin{figure}[t]
   \includegraphics[scale=0.3]{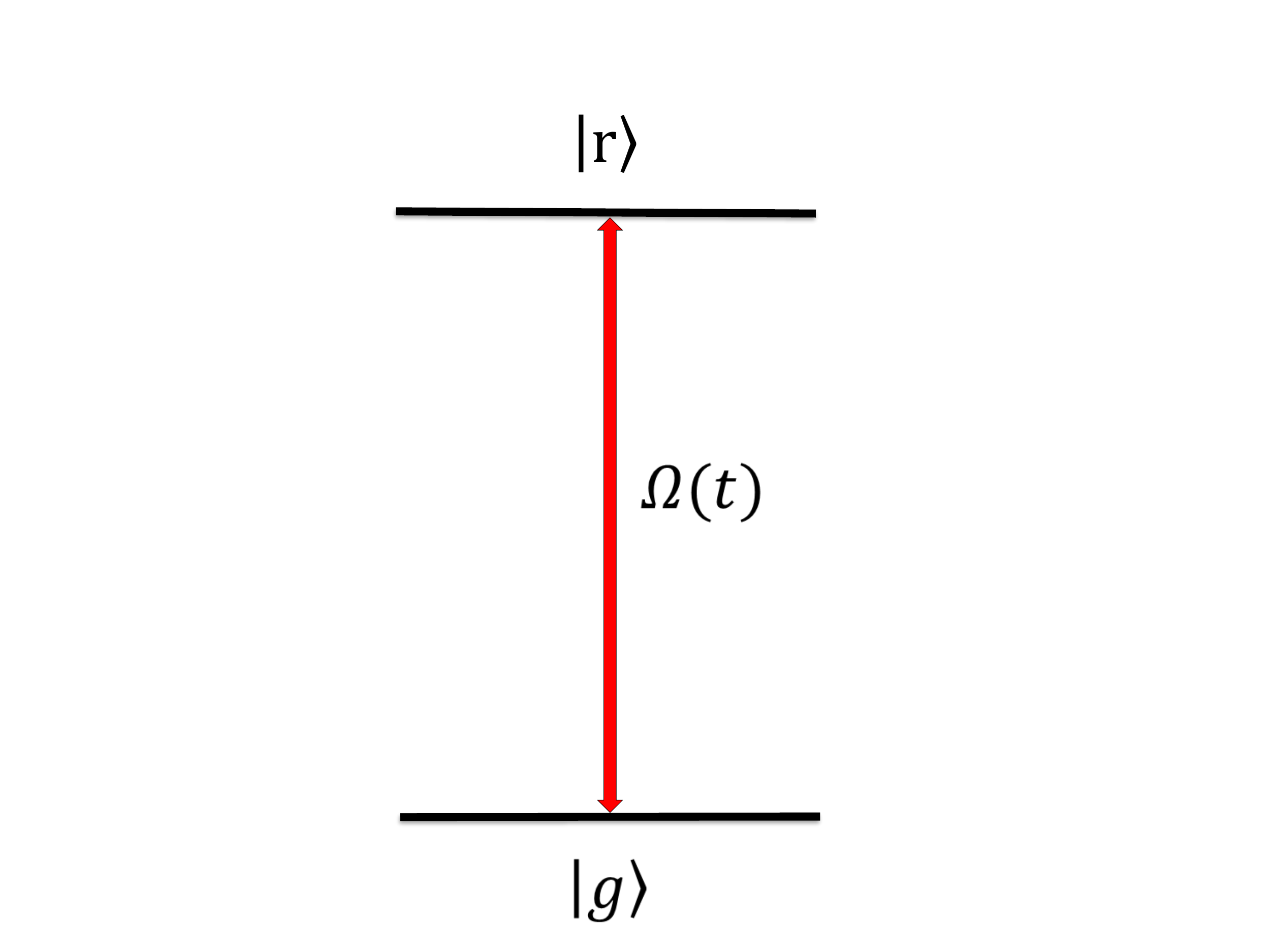}
   \caption{(Color online) Setup for two-level Rydberg atom. A stable ground states $|g\rangle$ is coupled to Rydberg state $|r\rangle$ by resonant laser pulse with Rabi frequency $\Omega(t)$.} \label{fig1}
\end{figure}

Consider a two-level Rydberg atom consisting of a stable ground state $|g\rangle$ and a long-lived Rydberg state $|r\rangle$, where the transition  $|g\rangle\leftrightarrow|r\rangle$ is driven  by resonant laser pulse with Rabi frequency $\Omega(t)$, as shown in Fig. \ref{fig1}.
In the rotating frame, by using the rotating wave approximation, the Hamiltonian describing the Rydberg atom interacting with the laser reads
\begin{align}
H(t)=\Omega(t)|g\rangle\langle r|+\mathrm{H.c.}, \label{hamiltonian}
\end{align}
where $\mathrm{H.c.}$ represents the Hermitian conjugate term.

Taking $\{|g\rangle,|r\rangle\}$ as the computational basis, we aim to realize an arbitrary one-qubit nonadiabatic geometric gate,
\begin{align}
U_{\boldsymbol{n}}(\gamma)=e^{i\gamma\boldsymbol{n\cdot\sigma}}. \label{operator}
\end{align}
Here, $\boldsymbol{\sigma}=(\sigma_{x},\sigma_{y},\sigma_{z})$ are the standard Pauli operators acting on computational basis $|g\rangle$ and $|r\rangle$, $\boldsymbol{n}=(\sin\theta\cos\varphi,\sin\theta\sin\varphi,\cos\theta)$ is an arbitrary unit vector, and $\gamma$ is an arbitrary phase with the geometric feature. $U_{\boldsymbol{n}}(\gamma)$ represents a rotational gate with the rotational axis $\boldsymbol{n}$ and rotational angle $\gamma$.

We now demonstrate how to realize such a gate by using the Hamiltonian described by Eq. (\ref{hamiltonian}). To this end, we take the Rabi frequency of the laser pulse as
\begin{align}
\Omega(t)=\begin{dcases}
~\Omega_{R}(t)e^{-i(\varphi-\frac{\pi}{2})}, & 0\leq t \leq \tau_1, \\
~\Omega_{R}(t)e^{-i(\gamma+\varphi+\frac{\pi}{2})}, & \tau_1 < t \leq \tau_2, \\
~\Omega_{R}(t)e^{-i(\varphi-\frac{\pi}{2})}, & \tau_2 < t \leq \tau, \\
\end{dcases}
\end{align}
where the amplitude parameter $\Omega_{R}(t)$ is time-dependent and  the phase parameter $\varphi$ is time-independent. Here,  intermediate time $\tau_1$,  $ \tau_2$ and  the total evolution time $\tau$ satisfy
\begin{align}
\int^{\tau_{1}}_{0}\Omega_{R}(t)dt=\frac{\theta}{2}, ~\int^{\tau_{2}}_{\tau_{1}}\Omega_{R}(t)dt&=\frac{\pi}{2}, ~\int^{\tau}_{\tau_{2}}\Omega_{R}(t)dt=\frac{\pi}{2}-\frac{\theta}{2}.\label{thetapai}
\end{align}
Then, the evolution operator can be expressed as
\begin{align}
U(t)=\begin{dcases}
~e^{-i\int^{t}_{0}H_{1}(t^{\prime})dt^{\prime}}, & 0\leq t \leq \tau_1, \\
~e^{-i\int^{t}_{\tau_1}H_{2}(t^{\prime})dt^{\prime}}U(\tau_1), & \tau_1 < t \leq \tau_2, \\
~e^{-i\int^{t}_{\tau_2}H_{1}(t^{\prime})dt^{\prime}}U(\tau_2), & \tau_2 < t \leq \tau, \\
\end{dcases}
\label{ut}
\end{align}
with
\begin{align}
H_{1}(t)&=\Omega_{R}(t)e^{-i(\varphi-\frac{\pi}{2})}|g\rangle\langle r|+\mathrm{H.c.},\notag\\
H_{2}(t)&=\Omega_{R}(t)e^{-i(\gamma+\varphi+\frac{\pi}{2})}|g\rangle\langle r|+\mathrm{H.c.}.\label{h12}
\end{align}
Specially, at the final time $t=\tau$, there is
\begin{align}
U(\tau)=e^{-i\int^{\tau}_{\tau_{2}}H_{1}(t)dt}e^{-i\int^{\tau_{2}}_{\tau_{1}}H_{2}(t)dt}
e^{-i\int^{\tau_{1}}_{0}H_{1}(t)dt}.\label{utau}
\end{align}
From Eqs. (\ref{thetapai}) and (\ref{h12}), we can obtain
\begin{align}
e^{-i\int^{\tau_{1}}_{0}H_{1}(t)dt}&=\cos\frac{\theta}{2}+\sin\frac{\theta}{2}\left(e^{-i\varphi}\ket{g}\bra{r}- e^{i\varphi}\ket{r}\bra{g}\right),\notag\\
e^{-i\int^{\tau_{2}}_{\tau_1}H_{2}(t)dt}&=-e^{-i(\varphi+\gamma)}\ket{g}\bra{r}+ e^{i(\varphi+\gamma)}\ket{r}\bra{g},\notag\\
e^{-i\int^{\tau}_{\tau_2}H_{1}(t)dt}&=\sin\frac{\theta}{2}+\cos\frac{\theta}{2}\left(e^{-i\varphi}\ket{g}\bra{r}- e^{i\varphi}\ket{r}\bra{g}\right).
\label{u123}
\end{align}
Substituting Eq. (\ref{u123}) into Eq. (\ref{utau}), we have, in the basis $\{\ket{g},\ket{r}\}$,
\begin{align}
U(\tau)=\cos\gamma+i\sin\gamma\left(
  \begin{array}{cc}
   \cos\theta&\sin\theta e^{-i\varphi }\\
   \sin\theta e^{i\varphi}& -\cos\theta
   \end{array}
\right).
\end{align}
It can be written as
\begin{align}
U(\tau)=U_{\boldsymbol{n}}(\gamma)=e^{i\gamma\boldsymbol{n\cdot\sigma}} \label{tong1}
\end{align}
with the aid of $\boldsymbol{n}=(\sin\theta\cos\varphi,\sin\theta\sin\varphi,\cos\theta)$, or equivalently
\begin{align}
U(\tau)=e^{i\gamma}\ket{d}\bra{d}+ e^{-i\gamma}\ket{b}\bra{b}, \label{tong1b}
\end{align}
where
\begin{align}
&|d\rangle=\cos\frac{\theta}{2}|g\rangle+\sin\frac{\theta}{2}e^{i\varphi}|r\rangle,\notag\\
&|b\rangle=\sin\frac{\theta}{2}e^{-i\varphi}|g\rangle-\cos\frac{\theta}{2}|r\rangle, \label{consition}
\end{align}
are the eigenstates of $\boldsymbol{n\cdot\sigma}$.

Equation (\ref{tong1}) as well as Eq. (\ref{tong1b})  indicates that the evolution operator $U(\tau)$ is exactly the rotational gate with the rotational axis $\boldsymbol{n}$ and rotational angle $\gamma$.

\begin{figure}[t]
   \includegraphics[scale=0.3]{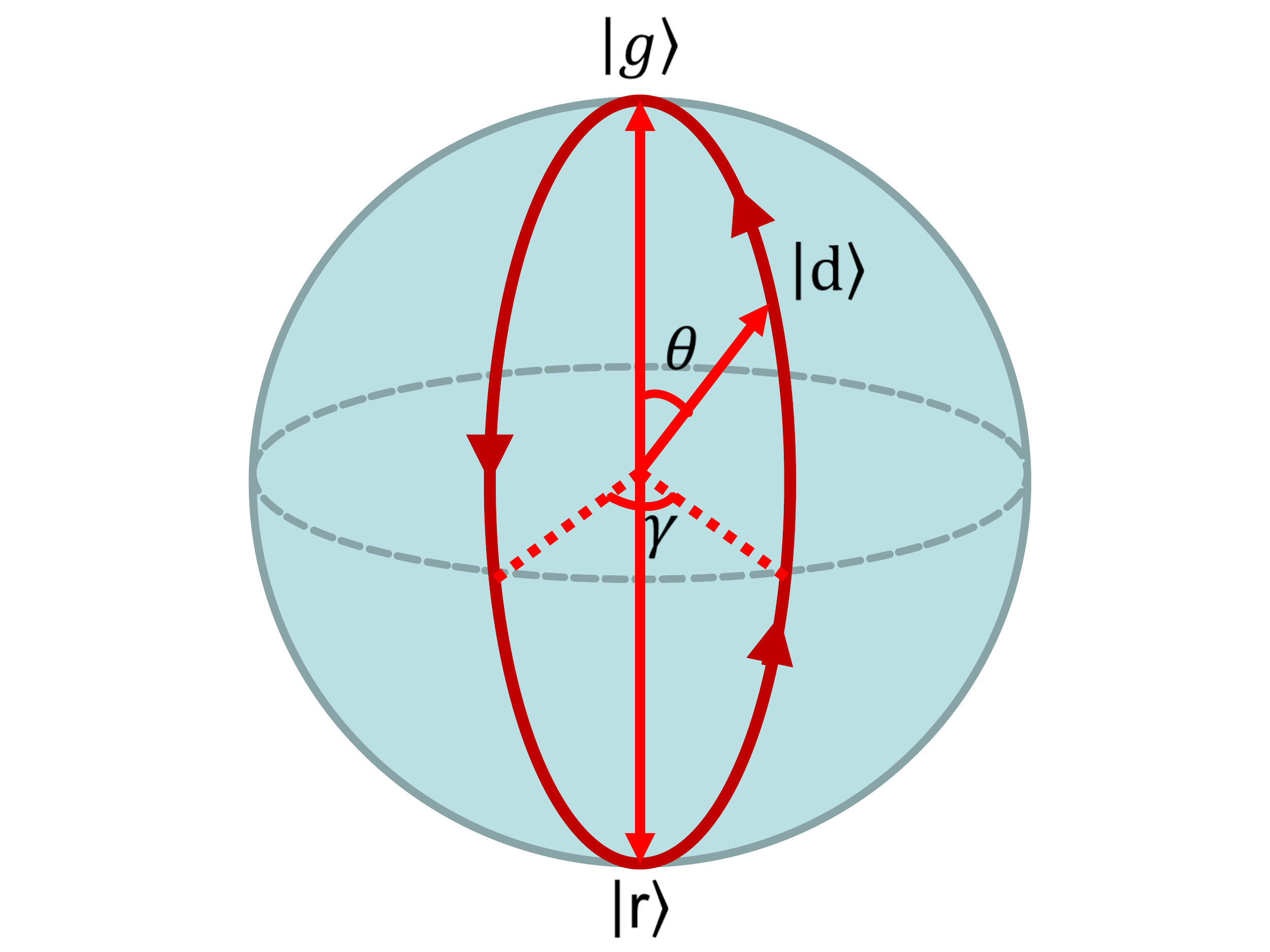}
   \caption{(Color online) Bloch sphere representation for the evolution of state $|d\rangle$. The state $|d\rangle$ undergoes cyclic evolution, such that $|d\rangle \xrightarrow{e^{-i\int^{t}_{0}H_{1}(t)dt}} |g\rangle   \xrightarrow{e^{-i\int^{t}_{\tau_{1}}H_{2}(t)dt}}  e^{i(\gamma+\varphi)}|r\rangle  \xrightarrow{e^{-i\int^{t}_{\tau_{2}}H_{1}(t)dt}} e^{i\gamma}|d\rangle$, and
    acquires a geometric phase $\gamma$. In this process, the dynamical phase is always zero, i.e.,
    $\langle d(t)|H_{1}(t)|d(t)\rangle=0$ $(0\leq t \leq \tau_1)$, $\langle d(t)|H_{2}(t)|d(t)\rangle=0$ $(\tau_1 < t \leq \tau_2)$, and $\langle d(t)|H_{1}(t)|d(t)\rangle=0$ $(\tau_2 < t \leq \tau)$, where $\ket{d(t)}=U(t)\ket{d}$ and $U(t)$ is defined in Eq. (\ref{ut}).} \label{fig2}
\end{figure}

To show $U(\tau)$ is a geometric gate, we examine the evolution of state $|d\rangle$. Clearly, the state $|d\rangle$ undergoes cyclic evolution,
\begin{align}
|d\rangle \xrightarrow{e^{-i\int^{t}_{0}H_{1}(t)dt}} |g\rangle   \xrightarrow{e^{-i\int^{t}_{\tau_{1}}H_{2}(t)dt}}  e^{i(\gamma+\varphi)}|r\rangle  \xrightarrow{e^{-i\int^{t}_{\tau_{2}}H_{1}(t)dt}} e^{i\gamma}|d\rangle.
\end{align}
It means that $\gamma$ is the total phase obtained by the state $|d\rangle$  undergoing cyclic evolution. The dynamic phase is always zero during the whole evolution, since the parallel transport condition is satisfied, i.e.,
\begin{align}
\langle d(t)|H_{1}(t)|d(t)\rangle&=0,~~~0\leq t \leq \tau_1,  \notag\\
\langle d(t)|H_{2}(t)|d(t)\rangle&=0,~~~\tau_1 < t \leq \tau_2, \notag\\
\langle d(t)|H_{1}(t)|d(t)\rangle&=0,~~~ \tau_2 < t \leq \tau,
\end{align}
where $\ket{d(t)}=U(t)\ket{d}$ and $U(t)$ is defined in Eq. (\ref{ut}).
Therefore, $\gamma$ is a purely geometric phase without containing any dynamic phase. Indeed, $\gamma$ is just equal to half of the solid angle enclosed by the orange-slice-shaped loop, shown in Fig. {\ref{fig2}}, which is the evolution path traced by $U(t)\ket{d}$ in the Bloch sphere. Similar discussion is available for state $\ket{b}$.

It is worth noting that an arbitrary one-qubit nonadiabatic geometric gate in our scheme is realized by a single orange-slice-shaped loop rather than combining multiple single orange-slice-shaped loops. Compared with the previous schemes of the one-qubit nonadiabatic geometric gates based on orange-slice-shaped loop \cite{Thomas2011,Mousolou2014PRA,Xu2014PRA,Xu2014}, in which a general one-qubit gate is realized by combining several special gates,  our scheme  minimizes the exposure time of gates to error sources but also keeps all the merits of the previous schemes. Besides, the implementation of our one-qubit gate is only based on a general two-level system driven by a laser pulse, which is not limited to a Rydberg atom but is applicable to any other two-level systems.

\section{Two-qubit gate}

We have shown that an arbitrary one-qubit nonadiabatic geometric gate can be obtained by addressing an individual Rydberg atom with laser pulse. To realize nonadiabatic geometric quantum computation, beside one-qubit gates, a nontrivial two-qubit gate is needed. We now demonstrate how to realize a nontrivial two-qubit nonadiabatic geometric gate by using the Rydberg-mediated interaction.

\begin{figure}[t]
   \includegraphics[scale=0.3]{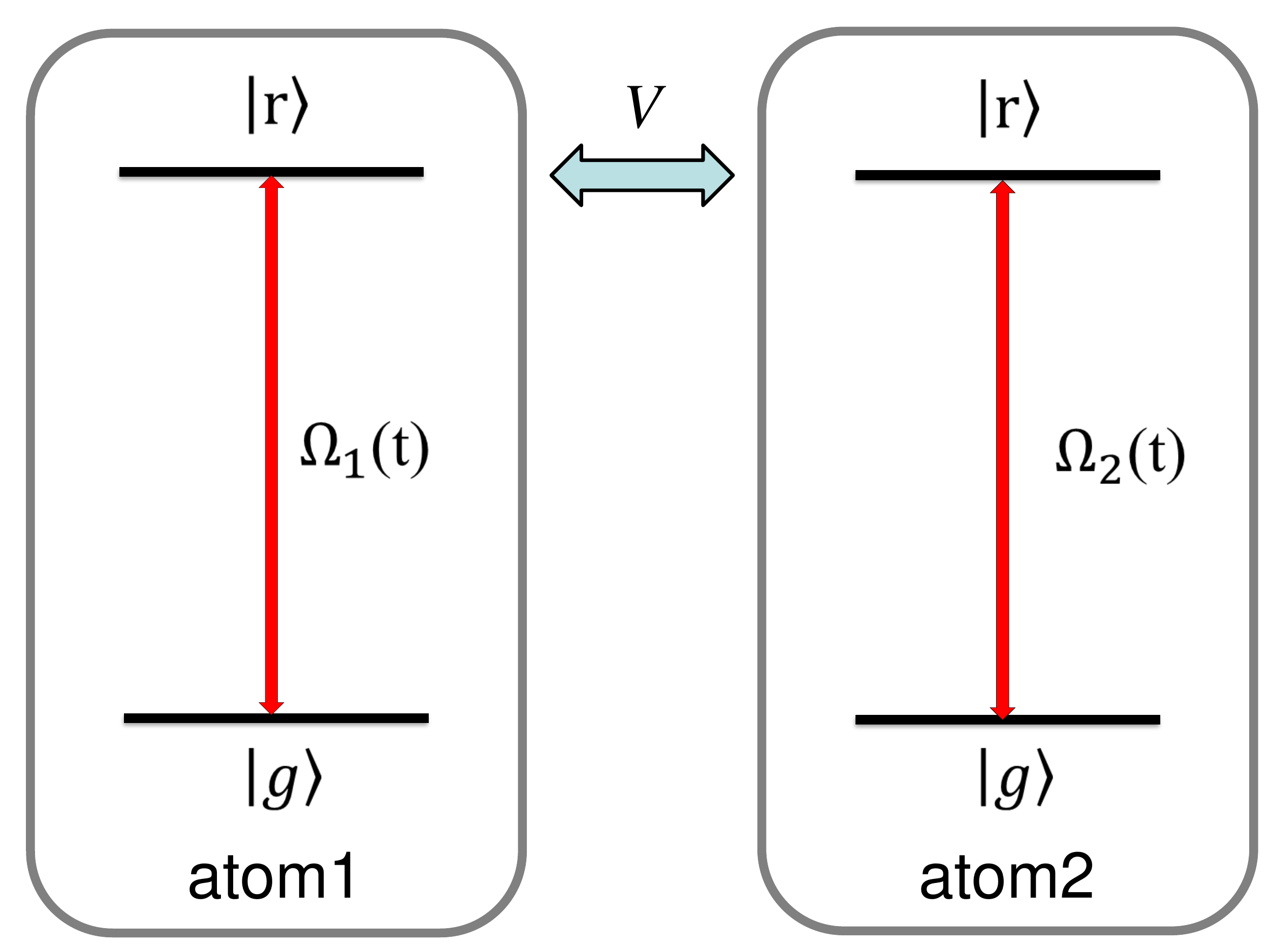}
   \caption{(Color online) Setup for two identical Rydberg atoms with Rydberg-mediated interaction. The interaction between Rydberg atoms is dipole-dipole or van der Waals force with strength $V$. For the individual Rydberg atoms, the transition $|g\rangle_{\alpha}\rightarrow|r\rangle_{\alpha}$ is facilitated by resonant laser pulse with Rabi frequency $\Omega_{\alpha}(t)$. } \label{fig3}
\end{figure}

Consider two two-level Rydberg atoms with Rydberg-mediated interaction $V$ arising from dipole-dipole or van der Waals force between high-lying Rydberg states, as shown in Fig. {\ref{fig3}}.
For each of the two Rydberg atoms, the transition $|g\rangle_{\alpha}\rightarrow|r\rangle_{\alpha}$ of the $\alpha$th Rydberg atom is driven by a resonant laser pulse with Rabi frequency $\Omega_\alpha(t)$, where $\alpha=1,2$. The computational space is taken as $S=\mathrm{Span}\{|gg\rangle,|gr\rangle,|rg\rangle,|rr\rangle\}$, where $|\mu\nu\rangle=|\mu\rangle_1|\nu\rangle_2$ with $\mu,\nu=g,r$. In the rotating frame, the Hamiltonian of the two-atom system reads
\begin{align}
H_{12}(t)=\mathcal{H}_1(t)\otimes I_2 +I_1\otimes \mathcal{H}_2(t)+V|rr\rangle\langle rr|, \label{hamiltonian1}
\end{align}
where
\begin{align}
\mathcal{H}_\alpha(t)=\Omega_\alpha(t)(|g\rangle_{\alpha}\langle r|+|r\rangle_{\alpha}\langle g|),
\end{align}
is the single-atom Hamiltonian describing the interaction between the $\alpha$th Rydberg atom and the laser pulse, and $I_{\alpha}$ denotes the identity operator acting on the $\alpha$th Rydberg atom. Here, we have neglected the rapidly oscillating counter-rotating terms in $\mathcal{H}_\alpha(t)$. In order to realize a nontrivial two-qubit nonadiabatic geometric gate, the Rabi frequencies of laser pulses are taken as $\Omega_{1}(t)=-\Omega_R(t)\cos(\phi/2)$ and $\Omega_{2}(t)=\Omega_R(t)\sin(\phi/2)$, where $\Omega_R(t)$ is time-dependent and $\phi$ is time-independent.
With the aid of these relations, we have,  by substituting $\mathcal{H}_1(t)$ and $\mathcal{H}_2(t)$ into Eq. (\ref{hamiltonian1}),
\begin{align}
H_{12}(t)=\Omega_R(t)\Big(|B\rangle\langle gg|-|B^{\prime}\rangle\langle rr|+\mathrm{H.c.}\Big)+V|rr\rangle\langle rr|,
\end{align}
where $|B\rangle =\sin(\phi/2)|gr\rangle-\cos(\phi/2)|rg\rangle$, $|B^{\prime}\rangle =\cos(\phi/2)|gr\rangle-\sin(\phi/2)|rg\rangle$. Further, we take a rotation transform $\mathcal{U}=\exp({-iV|rr\rangle\langle rr|t})$, and then the two-atom Hamiltonian can be recast as
\begin{align}
H_{\mathrm{rot}}(t)=\Omega_R(t)\left(|B\rangle\langle gg|-|B^{\prime}\rangle\langle rr|e^{-iVt}\right)+\mathrm{H.c.}. \label{hamiltonian3}
\end{align}
In Eq. (\ref{hamiltonian3}), state $|gg\rangle$ is resonantly coupled to state $|B\rangle$ with $\Omega_{R}(t)$, while $|rr\rangle$ is off-resonantly coupled to $|B^{\prime}\rangle$ due to the Rydberg-mediated interaction $V$.
If $V$, compared with $\Omega_{R}(t)$, is sufficiently large, i.e., $V\gg\Omega_R(t)$, the off-resonant terms are negligible. It means that the simultaneous excitation of two atoms from ground states to Rydberg states is inhibited.
In this case, the effective Hamiltonian reads
\begin{align}
H_{\mathrm{eff}}(t)=\Omega_R(t)\left(|B\rangle\langle gg|+|gg\rangle\langle B|\right). \label{hamiltonian4}
\end{align}

Since the Rydberg-mediated interaction $V$, being adjustable, can be easily tuned to a value much larger than $\Omega_R(t)$,  the above effective Hamiltonian is available. It is interesting to note that $H_{\mathrm{eff}}(t)$ has some desired features when it acts on the computational space. Note that the computational space $S=\mathrm{Span}\{|gg\rangle,|gr\rangle,|rg\rangle,|rr\rangle\}$ comprises of three subspaces $S_{1}=\mathrm{Span}\{|gg\rangle\}$, $S_2=\mathrm{Span}\{|gr\rangle,|rg\rangle\}$ and $S_3=\mathrm{Span}\{|rr\rangle\}$, where $S_2$ can be equivalently rewritten as  $S_2=\mathrm{Span}\{|B\rangle,|D\rangle\}$  with $|D\rangle =\cos(\phi/2)|gr\rangle+\sin(\phi/2)|rg\rangle)$. Subspace $S_1$ is coupled to subspace $S_{2}$ by $H_{\mathrm{eff}}(t)$, while $S_{3}$ remains unaffected by the Hamiltonian. Furthermore,  the state $|D\rangle$ in subspace $S_{2}$ is a dark state of the effective Hamiltonian, which is decoupled from other states. With the knowledge of these features of the Hamiltonian, we can construct a nontrivial two-qubit nonadiabatic geometric gate by carefully choosing the period of evolution.

The evolution operator reads
\begin{align}
U(t)=&e^{-i\int^{t}_{0}H_{\mathrm{eff}}(t^{\prime})dt^{\prime}}\notag\\
=&\cos\alpha_t(\ket{B}\bra{B}+\ket{gg}\bra{gg})-i\sin\alpha_t(\ket{B}\bra{gg}+\ket{gg}\bra{B})\notag\\
&+\ket{D}\bra{D}+\ket{rr}\bra{rr}
\label{u12}
\end{align}
where $\alpha_t=\int^{t}_{0}\Omega_R(t^{\prime})dt^{\prime}$. In the basis $\mathcal{S}=\mathrm{Span}\{|gg\rangle,|gr\rangle,|rg\rangle,|rr\rangle\}$, it can be expressed as
\begin{widetext}
\begin{align}
U(t)=\left(\begin{array}{cccc}
   \cos\alpha_t&-i\sin\alpha_t\sin\frac{\phi}{2}& i\sin\alpha_t\cos\frac{\phi}{2}&0 \\
 \ -i\sin\alpha_t\sin\frac{\phi}{2}&\cos^{2}\frac{\phi}{2}+\cos\alpha_t\sin^{2}\frac{\phi}{2}
 &\sin\frac{\phi}{2}\cos\frac{\phi}{2}(1-\cos\alpha_t)&0\\
    \ i\sin\alpha_t\cos\frac{\phi}{2}&\sin\frac{\phi}{2}\cos\frac{\phi}{2}(1-\cos\alpha_t)
    &\sin^{2}\frac{\phi}{2}+
    \cos\alpha_t\cos^{2}\frac{\phi}{2}&0 \\
    0&0&0&1
  \end{array}
\right).
\label{u12b}
\end{align}
\end{widetext}

If we let the evolution period $\tau$ satisfy
\begin{align}
\alpha_\tau=\int^{\tau}_{0}\Omega_R(t)dt=\pi, \label{thetapai1}
\end{align}
we have, from  Eq. (\ref{u12b}),
\begin{align}
U(\tau)=\left(
  \begin{array}{cccc}
   -1 & 0 & 0 & 0 \\
   0 & \cos\phi & \sin\phi & 0 \\
   0 & \sin\phi & -\cos\phi & 0 \\
   0 & 0 & 0 & 1 \\
  \end{array}
\right),
\end{align}
which provides a nontrivial two-qubit gate.

To show  $U(\tau)$ is a geometric one, we rewrite it as
\begin{align}
U(\tau)=U_{S_1}(\tau)\bigoplus U_{S_2}(\tau)\bigoplus U_{S_3}(\tau),
\end{align}
where  $U_{S_1}(\tau)=-1$, $U_{S_2}(\tau)=\left(\begin{array}{cc} \cos\phi & \sin\phi \\ \sin\phi & -\cos\phi  \end{array} \right),$ and $U_{S_3}(\tau)=1$. Here, $U_{S_1}(\tau)$,  $U_{S_2}(\tau)$ and $U_{S_3}$ act on the subspaces $S_1$, $S_2$, and $S_3$, respectively. Since $U(t)$ does not affect $S_3$, we only need to show that $U_{S_1}(\tau)$ is a geometric phase factor  and $U_{S_2}(\tau)$ is a holonomic matrix. Indeed, $U_{S_1}(\tau)$ is a geometric phase factor, since it is just the phase factor gained by the basis $\ket{gg}$ undergoing cyclic evolution and the evolution satisfies the parallel transport condition
\begin{align}
\bra{gg}U^\dag(t)H_{\mathrm{eff}}(t)U(t)\ket{gg}=0,
\end{align}
for $0\leq t \leq \tau$.   Further, one can verify that the following conditions,
\begin{align}
U^\dag(\tau)\left(\ket{gr}\bra{gr}+\ket{rg}\bra{rg}\right)U(\tau)=\ket{gr}\bra{gr}+\ket{rg}\bra{rg},
\end{align}
and
\begin{align}
&\bra{gr}U^\dag(t)H_{\mathrm{eff}}(t)U(t)\ket{gr}=\bra{rg}U^\dag(t)H_{\mathrm{eff}}(t)U(t)\ket{rg}=0,\notag\\
&\bra{gr}U^\dag(t)H_{\mathrm{eff}}(t)U(t)\ket{rg}=\bra{rg}U^\dag(t)H_{\mathrm{eff}}(t)U(t)\ket{gr}=0,
\end{align}
for $0\leq t \leq \tau$, are fulfilled. It means that $U_{S_2}(\tau)$ is indeed a holonomic matrix \cite{Zanardi,Sjoqvist2012,Xu2012}.

The above demonstrations show that a state initially in subspace $S_{1}$  will undergo cyclic evolution and acquire nonadiabatic geometric phase $\pi$, a state initially in subspace $\mathcal{S}_{2}$ will undergo cyclic evolution and acquires a non-Abelian geometric phase, and a state initially in subspace $\mathcal{S}_{3}$ will remain unchanged. Note that a general state is a superposition of three components in the three subspaces, and the component in each subspace undergoes cyclic evolution. While every component in the subspaces acquires a geometric phase, the  nontrivial two-qubit geometric gate is realized. It combines the ideas of Abelian and non-Abelian geometric phases in one single gate.

Compared with the previous optical system-based schemes of nonadiabatic geometric two-qubit gates \cite{Sjoqvist2012,Kim2008,Feng2007,Feng2009,Liang2014}, which are realized by weak-field coupling, our two-qubit nonadiabatic geometric gate is realized by using resonant laser pulses, which is much more effective in the practical manipulation due to the strong resonant coupling between Rydberg states and ground states. Besides, the controllable Rydberg-mediated interaction facilitate its manipulation too.

\section{Discussions}
We have obtained an arbitrary one-qubit gate and a nontrivial two-qubit gate comprising a
universal set of nonadiabatic geometric gates. Quantum computation based on such gates possesses the
following merits: robustness against control errors, speediness of nonadiabatic evolution, long
coherence time of qubit states, and realistically controllable interaction between qubits. It is the
purpose of developing Rydberg atom-based nonadiabatic geometric gates to realize fast quantum
computation with the merits of both geometric phases and Rydberg atoms.

Our scheme realizes unitary gates, i.e., operations where the decoherence has been neglected due
to the long coherence time of the excited Rydberg states. However, many experimentally feasible
Rydberg levels carry remnant decay, which determines the coherence time and may still affect the
quantum gates. Noting that the evolution time of the gates, which satisfies Eq. (\ref{thetapai}) or
(\ref{thetapai1}), is restricted by the amplitude parameter $\Omega_{R}(t)$, a natural question is
whether the coherence time can sufficiently match the time requirement of time-dependent
function $\Omega_{R}(t)$ of the driving fields' Rabi frequencies. That is, can we properly choose
the amplitude parameter $\Omega_{R}(t)$ such that $\tau < \tau_c$ is satisfied and
$\Omega_{R}(t)\ll V$ is experimentally allowed, where $\tau_c$ represents the lifetime of a
Rydberg state and $V$ is the Rydberg-mediated interaction? Here, we answer this question.
As it has been shown \cite{Saffman2005,Moller2008}, for alkali atoms, the lifetime of a Rydberg
state with principal quantum number $n>70$ can be larger than $100\mathcal{\mu}s$ and the
dipole-dipole interaction strength between two atoms are above $200\pi\mathrm{MHz}$ when
atoms are separated less than $5\mathcal{\mu}\mathrm{m}$. On the other hand, Eqs. (\ref{thetapai})
and (\ref{thetapai1}) indicate that $\int_0^{\tau} \Omega_{R}(t) dt \sim \pi$. This means that
the requirement, $\tau < \tau_c$ and $\Omega_{R}(t)\ll V$, can be indeed satisfied  by properly
choosing $\Omega_{R}(t)$. For example, the amplitude parameter can be taken as $\Omega_{R}
\sim 5\pi\mathrm{MHz}$, which is sufficiently small compared with the interaction
$V\sim 200\pi\mathrm{MHz}$ and is experimentally allowed \cite{Saffman2005}, and the
corresponding evolution time of quantum gates is $\tau=0.2\mathcal{\mu}s$, which is sufficiently
short compared to the lifetime $\geq100\mathcal{\mu}s$. Therefore, the effective Hamiltonian
in Eq. (\ref{hamiltonian4}) is available. Besides, it should be noted that to implement our scheme
experimentally, the conditions expressed in Eqs. (\ref{thetapai}) and (\ref{thetapai1}) must be
fulfilled.

It is worth noting that Rydberg atoms have been used to realize photonic pulse-based quantum gates in previous works  \cite{Barato2014,He2014,Khazali2015,Das2016}, where the computational qubits are encoded by photons and the Rydberg atomic ensemble only acts as mediator. By converting photonic pulses into Rydberg excitations, the logic operations between pulses can be directly performed. Our scheme is different from the photonic pulse-based scenario. In our scheme, the computational qubits are encoded into the stable ground states of Rydberg atoms and the long-lived Rydberg states, and the quantum gates, based on nonadiabatic geometric phases, have the merits of both high-speed implementation and robustness against control errors. While photonic qubits in the Rydberg ensemble have the advantage that they can be directly read out, the operation in our scheme must be read out by some other means to convert it to flying qubits for the next  operation.

In passing, we point out that although the quantum gates in the present paper are demonstrated based on individual Rydberg atoms, our scheme can be extended to the mesoscopic atomic ensemble with the qubits encoded by collective Rydberg states of the atomic ensemble.

\section{Conclusion}
In conclusion, we have proposed a scheme to implement nonadiabatic geometric quantum computation with Rydberg atoms. By encoding the computational basis into the stable ground state and long-lived Rydberg state, a universal set of nonadiabatic geometric gates, consisting of an arbitrary one-qubit gate and a nontrivial two-qubit gate, is realized. The arbitrary one-qubit gate is performed by addressing an individual Rydberg atom with resonant laser pulses while the nontrivial two-qubit gate is performed with the aid of Rydberg blockade regime.

The one-qubit gate in our scheme is realized via a single orange-slice-shaped loop. Compared with all the previous one-qubit nonadiabatic geometric gates based on orange-slice-shaped loop, in which a general one-qubit gate is realized via combining several special gates, our scheme  minimizes the exposure time of gates to error sources but also keeps all the merits of the previous schemes. The two-qubit gate in our scheme is realized by using resonant laser pulses. Compared with all the previous optical system-based schemes of nonadiabatic geometric two-qubit gates, which are realized by weak-field coupling, our two-qubit nonadiabatic geometric gate is much more effective in the practical manipulation.
Our scheme combines the robustness of nonadiabatic geometric gates with the merits of Rydberg atoms, and thereby provides a promising approach to high-fidelity quantum computation.

\begin{acknowledgments}
P.Z.Z. acknowledges support from the National Natural Science Foundation of China though Grant No. 11575101. G.F.X. acknowledges support from the National Natural Science Foundation of China through Grant No. 11605104, from the Future Project for Young Scholars of Shandong University through Grant No. 2016WLJH21, and from the Carl Tryggers Stiftelse (CTS) through Grant No. 14:441. E.S. acknowledges financial support from the Swedish Research Council (VR) through Grant No. D0413201. D.M.T. acknowledges support from the National Basic Research Program of China through Grant No. 2015CB921004.
\end{acknowledgments}

\end{document}